\documentclass[aps,twocolumn,prl,showpacs,nofootinbib]{revtex4}
\usepackage{amsmath}
\usepackage{graphicx}
\usepackage{dcolumn}
\usepackage{bm}
\usepackage{amssymb}
\usepackage{latexsym}

\def\setC{\mathbb{C}}

\def\setR{\mathbb{R}}

\newcommand{\dd}{\mathrm{d}}
\newcommand{\ee}{\mathrm{e}}

\newcommand{\ie}{\textsl{i.e.~}}

\newcommand{\apriori}{\textsl{a priori~}}

\newcommand{\Ups}{\Upsilon}

\newcommand{\mP}{m_{_\mathrm{Pl}}}
\newcommand{\GReCO}{${\cal G}\setR\varepsilon\setC{\cal O}$}

\def\spose#1{\hbox to 0pt{#1\hss}}
\def\lta{\mathrel{\spose{\lower 3pt\hbox{$\mathchar"218$}}
     \raise 2.0pt\hbox{$\mathchar"13C$}}}
\def\gta{\mathrel{\spose{\lower 3pt\hbox{$\mathchar"218$}}
     \raise 2.0pt\hbox{$\mathchar"13E$}}}

\newcommand{\zBST}{\zeta_{_{\mathrm{BST}}}}

\newcommand{\Hu}{{\cal H}} 

\begin{document}

\title{On the ``Causality Argument'' in Bouncing Cosmologies}

\author{J\'er\^ome Martin} 
\email{jmartin@iap.fr}
\affiliation{Institut d'Astrophysique de
Paris, \GReCO, FRE 2435-CNRS, 98bis boulevard Arago, 75014 Paris,
France}

\author{Patrick Peter}
\email{peter@iap.fr}
\affiliation{Institut d'Astrophysique de
Paris, \GReCO, FRE 2435-CNRS, 98bis boulevard Arago, 75014 Paris,
France}

\date{November 3$^\mathrm{rd}$, 2003}

\begin{abstract}
We exhibit a situation in which cosmological perturbations of
astrophysical relevance propagating through a bounce are affected in a
scale-dependent way. Involving only the evolution of a scalar field in
a closed universe described by general relativity, the model is
consistent with causality. Such a specific counter-example leads to
the conclusion that imposing causality is not sufficient to determine
the spectrum of perturbations after a bounce provided it is known
before. We discuss consequences of this result for string motivated
scenarios.
\end{abstract}

\maketitle

It was recently acknowledged that perhaps the most fundamental key
questions of string~\cite{string} or M-theory could be addressed in
the framework of time-dependent cosmological
background~\cite{banks}. Conversely, it was found that this new
physics may imply new solutions for the dynamics of the early
Universe. It is the case in particular of the pre big-bang
paradigm~\cite{PBB}, or the more controversial~\cite{noekp} ekpyrotic
scenario~\cite{ekp}. In both these last two models, the effective four
dimensional theory presents a transition between a contracting and an
expanding phase, \ie a bounce~\cite{bounce,GT}.

In general, both the contracting and expanding phases can be described
by well controlled low-energy physics for which high energy
corrections (the $\alpha'$ terms~\cite{string} in string theory for
instance) are negligible. As a consequence, before (and after) the
bounce takes place, it is possible to calculate unambiguously the
spectrum of gravitational perturbations. On the contrary, the bounce
itself demands full knowledge of such corrections which are, in
practice, either difficult to implement~\cite{addR} or simply
unknown. This is why propagating perturbations through the bounce
represents a technical challenge but is of uttermost importance, in
order to make contact with observational cosmology.

The impossibility to know precisely the dynamics of the bouncing phase
has led to postulate that its effect would reduce at most to a scale
independent modification of the overall amplitude~\cite{PBB,DV}. As a
matter of fact, both during the radiation to matter~\cite{mfb}
transition and through the preheating~\cite{FB}, which are two other
examples of short duration cosmological transitions, the spectrum is
propagated in a scale-independent way. The purpose of this letter is
to re-examine this assumption in the context of bouncing universes. We
show, by means of a specific counter-example, that such a cosmological
transition can affect large wavelengths in a scale-dependent way and
this without violating causality. Thus, this last argument cannot be
invoked to justify the abovementionned postulate.

The simplest way to study a nonsingular bounce is to consider general
relativistic Friedmann-Lema\^{\i}tre-Robertson-Walker (FLRW) models
with closed spatial sections and a scalar field~\cite{GT}.  We choose
$\eta =0$ to be the conformal time at which the bounce occurs. Then,
without loss of generality, the FLRW scale factor in the vicinity of
the bounce (\ie for $\eta\ll 1$) can be described by a power series
expansion in $\eta$ with one free parameter for each term of the
series. We choose those such that the scale factor can be approximated
by
\begin{equation}
\frac{a(\eta )}{a_0}\simeq 1+\frac{1}{2}\left(\frac{\eta }{\eta _0}\right)^2
+\delta \left(\frac{\eta }{\eta _0}\right)^3+\frac{5}{24}(1+\xi )
\left(\frac{\eta }{\eta _0}\right)^4 .
\label{aseries}
\end{equation}
The first free parameter $a_0$ represents the value of the scale
factor at the bounce. The second free parameter, namely $\eta _0$,
gives the typical timescale of the bounce (the physical duration of
the bounce is $t_0 = a_0 \eta_0$) and determines the corresponding de
Sitter-like tangent model~\cite{MP}, the deviation from which being
measured by the third and fourth free parameters $\delta $ and $\xi$,
hence the non intuitive coefficients in Eq.~(\ref{aseries}); they are,
in some sense similar to the slow-roll parameters of inflationary
cosmology although they are not restricted to small values. It has
been shown in Ref.~\cite{MP} that one should restrict attention to
$\-11/5 \leq \xi \lta -0.1$. If $\delta =0$, the bounce is
symmetric. For $\vert \eta /\eta _0\vert \gg 1$, the above model is
assumed to be connected to other cosmological epochs. Note also that
adding more terms in the series (\ref{aseries}) does not change the
following argument~\cite{MP}.

The free parameter $\eta _0$ is directly connected to the null energy
condition (NEC) $\rho + p \geq 0$, where $\rho $ and $p$ are
respectively the energy density and the pressure. Indeed, at the
bounce, Einstein equations imply that
\begin{equation}
\lim_{\eta\to 0}\left(\rho +p\right)=2\frac{\Ups}{a_0^2},
\label{rhopp}
\end{equation}
where the parameter $\Ups$ is defined by the following expression
\begin{equation}
\Ups \equiv 1 - \frac{1}{\eta _0^2}\simeq 
\frac{\varphi _0'^2}{\mP^2},
\end{equation}
where $\varphi$ is the scalar field driving the bounce, behaving as
$\varphi=\varphi_0+\varphi_0' \eta +\cdots$ (a prime denotes a
derivative with respect to conformal time) around the bounce, and
$\mP$ is the Planck mass. Combining both equations, we see that the
NEC is satisfied at the bounce provided $\vert \eta _0\vert \ge 1$,
the limiting case $\eta _0=1$ corresponding to the vacuum equation of
state $\rho =-p$. The previous considerations demonstrate that the
bounce cannot be made arbitrarily short if one wants to preserve the
NEC. Under this last condition, the short bounce limit is not $\eta
_0\rightarrow 0$ but rather $\eta _0\rightarrow 1$, \ie $\Ups
\rightarrow 0$. Note that since $\Ups$ is related to the kinetic
energy density of the scalar field to the Planck density, one expects
$\Ups \ll 1$.

Let us now turn to the study of the cosmological perturbations around
the previous model (see Ref.~\cite{mfb}). The gravitational
fluctuations are characterized by the curvature perturbations on zero
shear hypersurfaces, \ie the Bardeen potential $\Phi$, whose master
equation of motion reduces to that of a parametric oscillator given by
\begin{equation}
\label{eomu}
u''+\left[n\left(n+2\right)-V_u\left(\eta \right)\right]u=0,
\end{equation}
where the factor $n(n+2)$ arises from the eigenvalue of the
Laplace-Beltrami operator on the closed spatial sections (hence $n$ is
an integer). In the above equation, we have introduced a new
gauge-invariant quantity, $u$, related to $\Phi$ by (see
Ref.~\cite{MP} for details) $\Phi \equiv \sqrt{3\kappa}\Hu u
/(2a^2\theta)$,
where $\kappa =8\pi /\mP^2$ and ${\cal H}\equiv a'/a$. The
quantity $\theta $ is defined by $\theta^{-2} \equiv a^2
\left(1+p/\rho\right)\left( 1+\Hu^{-2}\right)$. This quantity only
depends on the scale factor and its derivatives (up to the fourth
order).

It is well known that only those modes having $n>1$ are not gauge
modes. A crucial point is that the values of $n$ of astrophysical
interest today are such that $n\gg 1$ (e.g. for $\Omega_\mathrm{now}
\sim 1.01$, $60<n<6\times 10^6$). They are related to the more usual
wavenumber $k^2$, commonly used in inflationary cosmology, by the
relation $k^2=n(n+2)(\Omega -1)$ which shows that, in this last
context, $k^2\ll 1$ (since $\Omega \simeq 1$ at the end of inflation
with a very high accuracy). The effective potential $V_u(\eta )$ in
Eq.~(\ref{eomu}) can be expressed as
\begin{equation}
\label{potu}
V_u(\eta )=\frac{\theta ''}{\theta }+3(1-c_{_{\rm S}}^2),
\end{equation}
where $c_{_{\rm S}}^2\equiv p'/\rho '$ which, in some regimes, can be
interpreted as the sound velocity. Contrary to the flat case, the
effective potential cannot be cast into the form of the second
derivative of a function over that function.

The fact that the effective potential only depends on the scale factor
and its derivatives means that it can be calculated for a general
bounce given by Eq.~(\ref{aseries}). Far from the bounce, but still
with $\eta<\eta_0$, we have $V_u(\eta )\simeq 4$. In the vicinity of
the bounce and in the limit of a short bounce satisfying the NEC, one
has
\begin{equation}
\lim_{\eta\to 0}V_u=\frac{27\delta ^2}{\Ups ^2}
-\frac{81\delta ^2}{\Ups }+\frac{5\xi }{2\Ups }+{\cal O}(\Ups ^0).
\end{equation}
Let us notice that $V_u(\eta =0)$ is an extremum of the potential only
in the case $\delta =0$. We see that the height of the effective
potential diverges in the limit $\Ups \rightarrow 0$. Therefore, in
this limit, the bounce will necessarily affect the propagation of all
the Fourier modes, regardless of the values of the parameters $a_0$,
$\delta $ or $\xi $. In other words, the spectrum of the fluctuations
cannot \apriori propagate through the bounce without being changed in
the NEC-preserved short time limit. This conclusion is generic and
does not depend on the details of the model.

Is this compatible with causality? Let us first recall that causality
relies on the concept of horizon which, requiring an integration over
time $d_{_{\rm H}}(t)=a(t)\int _{t_{\rm i}}^t a^{-1}(\tau )\dd\tau$,
$t_{\rm i}$ being the initial time (which should be
$t_\mathrm{i}=-\infty $ in the case of nonsingular cosmology), is a
global quantity. Assuming from the outset an homogeneous background is
problematic whenever the implied horizon is finite since in this case
there exists scales larger than $d_{_{\rm H}}$; somehow, imposing any
condition on these scales is acausal. In other words, $d_{_{\rm H}}$
fixes the scale limit below which the theory is physically
meaningfull. Technically, this means that whenever a homogeneous
background is assumed, one should proceed as follows: solve,
mathematically, the perturbation equations for all modes $\lambda$,
and then implement causality by restricting attention to those
satisfying $\lambda \leq d_{_{\rm H}}$. To be able to decide whether
this condition is met, one needs to embed the local bounce transition
into a complete global model providing $d_{_{\rm H}}$. In conclusion,
without knowledge of the function $a(t)$ for all $t$, there can be no
fundamental principle which would preclude large scales to be
spectrally affected by a local effective potential (\ref{eomu}). (Note
that bounces are usually implemented in order to regularize the
singularity, leading to a geodesically complete universe, and hence an
infinite horizon.)

Another possible source of confusion is the often made identification
of the Hubble scale $\ell _{_{\rm H}}=a^2/a'$, a local quantity, with
the horizon $d_{_{\rm H}}$. Once a given scale $\lambda$ is inside the
horizon at some time $t=t_0$, it remains so for any time $t>t_0$
because the ratio of the horizon to the physical scale at time $t$ is
\begin{equation}
\frac{d_{_{\rm H}}}{\lambda }=\frac{k}{2\pi } \int _{t_{\rm
i}}^{t_0}\frac{{\rm d}\tau }{a(\tau )} + \frac{k}{2\pi } \int
_{t_0}^{t}\frac{{\rm d}\tau }{a(\tau )},
\end{equation}
where $k$ is the comoving wavenumber of the scale under
consideration. The first term is by assumption greater than unity and
the second one is positive definite, hence the statement. By the same
token, any scale which is outside the horizon will eventually enter it
later (unless a singularity develops before). In constrast, as is
well-known in the inflationary situation, a given scale can either
exit or enter the Hubble radius, which can be done because
$\lambda\geq\ell _{_{\rm H}}$ is physical, contrary to $\lambda\geq
d_{_{\rm H}}$. In the case of a bouncing universe, the Hubble radius
diverging at the bounce, all scales are, at some stage,
sub-Hubble. For flat spatial sections, a super-Hubble scale also means
that the mode is below the effective potential of the
perturbation. This is no longer the case in the curved bouncing model
for which the mode can be sub-Hubble although potential dominated, as
was discussed in Ref.~\cite{MP}.

\begin{figure}[h]
\hskip-2mm\includegraphics[width=9cm]{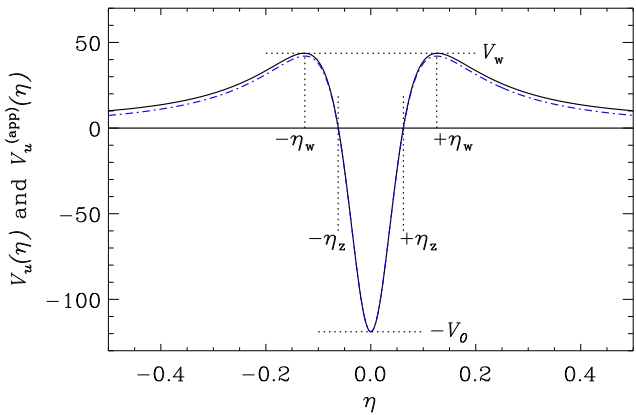}
\caption{The effective potential $V_u(\eta)$ for the perturbation
variable $u(\eta)$ as obtained by using the quartic expansion of the
scale factor. The values $\eta_0=1.01$, $\delta=0$, and $\xi=-2/5$
have been used to derive this plot. The almost undistinguishable
dot-dashed curve represents a rational approximation to this potential
which is valid in the vicinity of the bounce and that is used in
Ref.~\cite{MP}. Clearly, the potential is well-behaved at all
relevant times, and so is the corresponding variable $u$.}
\label{dSpot}
\end{figure}

We now calculate explicitly the effect of a symmetric bounce ($\delta
=0$) on the evolution of the cosmological perturbations. In this case,
it turns out~\cite{MP} that the effective potential can be expressed
as the ratio of two polynomials of order $24$; it is represented in
Fig.~\ref{dSpot}.
Obviously, for such a complicated potential, the equation of motion
(\ref{eomu}) cannot be solved analytically. However, in the case we
are interested in, namely $\Ups \ll 1$, the potential has a simple
behavior, which can be studied simply by retaining only the smallest
order terms in $\Ups$.

In the limit where $\Ups$ goes to zero, the extremum of the potential
$V_0$ diverges while its width, characterized by $\eta _{\rm z}$,
shrinks to zero, namely
\begin{equation}
V_0=-\frac{5\xi }{2\Ups }+{\cal O}(\Ups^0), \quad \eta _{\rm
z}=\sqrt{-\frac{\Ups}{5\xi }}+{\cal O}(\Ups ^{3/2}).
\end{equation}
This observation is confirmed by a study of the wings in
Fig.~\ref{dSpot} whose height and position are found to be
\begin{equation}
V_{\rm w}=-\frac{5\xi }{6\Ups }+{\cal O}(\Ups^0), \quad \eta _{\rm
w}=\sqrt{-\frac{4\Ups}{5\xi }}+{\cal O}(\Ups ^{3/2}),
\end{equation}
and exhibit therefore a similar behavior in the NEC-preserved short
time limit. The previous properties suggest that we deal with a
distributional effective potential, and indeed, a more careful
analysis reveals that~\cite{MP}
\begin{equation}
V_u(\eta )=-C_{\Ups}\Delta _{\Ups}(\eta ),
\end{equation}
where the constant $C_{\Ups}$ is given by $C_{\Ups}\equiv [-5\pi
^2\xi/(8\Ups) ]^{1/2}$ and where the function $\Delta _{\Ups }(\eta )$
is a representation of the Dirac $\delta $-function, \ie $ \lim _{\Ups
\rightarrow 0}\Delta _{\Ups }(\eta )=\delta (\eta )$. Note that even
though the potential appears singular, this is nothing but a
computational trick allowing an easy derivation of the resulting
spectrum (recall that in realistic situations, the parameter $\Ups$
must remain small but nonvanishing~\cite{MP}). The equation of motion
(\ref{eomu}) of the quantity $u$ can now be written as
\begin{equation}
u''+[n(n+2)+C_{\Ups }\delta (\eta )]u=0,
\end{equation}
which is nothing but a Schr\"odinger-like equation for a
distributional potential. Right before (superscript $<$) and after
(superscript $>$) the point $\eta=0$, but still within the bounce
epoch, a Fourier mode does not interact with the barrier since the
potential vanishes and the solutions are just linear combinations of
plane waves $\ee^{i\sqrt{n(n+2)}\eta }$ and
$\ee^{-i\sqrt{n(n+2)}\eta}$ with coefficients $A^{>}$, $B^{>}$ and
$A^{<}$, $B^{<}$ before and after the bounce respectively.

In order to calculate what is the spectrum after the bounce, being
given some initial conditions before the bounce, one must apply
junction conditions. In the case at hand, the matching conditions are
$[u]\equiv u(0^+) - u(0^-)=0$ and $[u']=-C_{\Ups }u(0)$, the last one
coming from an integration of the equation of motion in a thin shell
around $\eta =0$. This reduces to
\begin{eqnarray}
\label{junc1}
A^{>}+B^{>} &=& A^{<}+B^{<},
\\
\label{junc2}
A^{>}-B^{>} &=& A^{<}-B^{<} -\frac{C_{\Ups }(A^{<} +
B^{<}) }{i\sqrt{n(n+2)}}.
\end{eqnarray}
In the limit $\Ups \rightarrow 0$, the constant $C_{\Ups}$ diverges
and therefore the second term in Eq.~(\ref{junc2}) is
dominant. Straightforward algebraic manipulation allows us to
determine the transfer matrix defined by~\cite{DV}
\begin{equation}
\begin{pmatrix}
A^{>} \cr
B^{>}
\end{pmatrix}
=T_u
\begin{pmatrix}
A^{<} \cr
B^{<}
\end{pmatrix}, 
\end{equation}
and we obtain the following expression
\begin{equation}
\label{Tudelta}
T_u=-i\sqrt{\frac{-5\pi ^2\xi }{32n(n+2)}}
\begin{pmatrix}
1 & 1 \cr
-1 & -1
\end{pmatrix}
\frac{1}{\Ups ^{1/2}}.
\end{equation}
Some comments are in order at that point. First, one sees that, as
discussed above, the transfer matrix depends on the wavenumber. Our
calculation permits to actually predict accurately what the dependence
is: $\propto [n(n+2)]^{-1/2}$. Moreover, the calculation also predicts
the dependence of the transfer matrix on the parameter $\xi$ (except
in the limit $\xi\to 0$ for which a different calculation must be
done~\cite{MP}). A point worth mentioning is that the overall
amplitude diverges as $\Ups \rightarrow 0$. Since $u$ is just a
mathematically convenient variable, this is not necessarily
problematic. Indeed, using the relation between $u$ and the Bardeen
potential $\Phi$, and the fact that, at the bounce, Eq.~(\ref{rhopp})
holds, one can show that the spectrum of the Bardeen potential is
perfectly finite after the bounce, even in the limit $\Ups \rightarrow
0$. Note also that the fact that relevant scales may be larger than
the duration of the bounce itself does not preclude them to be
affected by the transition. Finally, it is worth mentioning that the
above result has been recovered in Ref.~\cite{MP} using a different
method, including the numerical factor in the overall amplitude.

In more standard situations, even though the amplitude of $\Phi$ might
change, the curvature perturbation on uniform density hypersurfaces,
\ie the quantity called $\zBST$~\cite{WMLL}, does not. Indeed, it
satisfies~\cite{MS}
\begin{equation}
\frac{\zBST'}{\Hu} \propto \left[
\frac{1}{3} \left(\frac{k}{\Hu}\right)^2 \left(
\frac{\Phi'}{\Hu}+\Phi\right) + \frac{\kappa a^2}{2\Hu^2}\delta
p_\mathrm{nad} \right],\label{zBST'}
\end{equation}
which was been shown~\cite{WMLL} to hold independently of the
gravitational field equations. Eq.~(\ref{zBST'}) implies that $\zBST$
is conserved under the conditions that there is no entropy
perturbation ($\delta p_\mathrm{nad}=0$), the decaying mode of $\Phi$
is neglected, and the scales are super-Hubble ($k\ll \Hu$).  Since
$\Hu=0$ at the bounce, the Hubble radius is larger than any relevant
scale during a finite interval around the bounce, \ie cosmological
scales are not large in the super-Hubble sense. Through the bounce,
moreover, the notion of decaying and growing modes is irrelevant. Two
conditions out of three being violated, $\zBST$ has no reason to be
conserved, and hence is not convenient for describing a bouncing
transition. One should note, additionnally, that since the spectrum of
$\Phi$ is altered, there is no reason why $\zBST$ should not be also
spectrally distorted through a bounce.

In this letter, we have demonstrated that there is no reason to
believe that the spectrum of large scale cosmological fluctuations is
not affected by a short duration bounce, although this is not
necessarily the case (one could choose for instance $\eta _0\gg 1$,
however hard this is to reconcile with the field theoretical
treatment~\cite{MP} or have a ``slow-roll'' kind of bounce with
$\eta_0=1$~\cite{GT}). We have shown that this occurs when one
approaches the NEC violation and we have also demonstrated that this
effect does not violate causality. This result may find important
applications: although the calculation discussed here is based on
general relativity, there is no reason why causality should act
differently in the framework of, say, string theory. In string
motivated cosmological scenarios (as for instance in the pre-big bang
paradigm~\cite{PBB} or in the ekpyrotic case~\cite{ekp}), the
calculation of the power spectrum of cosmological fluctuations is done
in the contracting phase and the predictions relevant for
observational purposes then stems from the assumption that for
sufficiently large scales, perturbations are essentially not affected
by the bounce. Our result indicates that this assumption is far from
trivial and may challenge the conclusions reached so far in the
literature.

We wish to thank R.~Brandenberger, A.~Buonanno and particularly
F.~Finelli and D.~J.~Schwarz for enlightening discussions over many
points discussed in this paper.

\end{document}